\documentstyle[twocolumn,aps,epsf]{revtex}
\newcommand{\avg}[1]{\langle{#1}\rangle}

\newcommand{\beq}{\begin{equation}}
\newcommand{\eeq}{\end{equation}}
\newcommand{\beqar}{\begin{eqnarray}}
\newcommand{\eeqar}{\end{eqnarray}}

\begin{document}
\draft
\wideabs
{
\title{Extracting Hidden Information from Knowledge Networks}
\author{Sergei Maslov$^{(1,2)}$, Yi-Cheng Zhang$^2$}

\address{$^1$ Department of Physics, Brookhaven National Laboratory,
Upton, New York 11973, USA}
\address{$^2$ Institut de Physique Th\'{e}orique, Universit\'{e} de
Fribourg, CH-1700, Fribourg, Switzerland}

\date{\today}
\maketitle
\begin{abstract}
We develop a method allowing us to reconstruct individual tastes
of customers from a sparsely connected network of their opinions
on products, services, or each other. Two distinct phase transitions
occur as the density of edges in this network is increased: above
the first - macroscopic prediction of tastes becomes possible, while
above the second - all unknown opinions can be uniquely reconstructed.
We illustrate our ideas using a simple Gaussian model, which we
study using both field-theoretical methods and numerical simulations.
We point out a potential relevance of our approach
to the field of bioinformatics.

\end{abstract}
\pacs{89.75.Hc, 89.20.Hh, 87.15.Kg, 89.65.Gh, 05.40.-a}
}
\narrowtext

Mainstream economics maintains that human tastes reflected in
consumer preferences are sovereign, i.e not subject to discussion
or study. It postulates that consumer's choice of products or
services is the outcome of a complete and thorough optimization
among all possible options, and, therefore, his/her satisfaction
cannot be further improved. Such a doctrine, though often
challenged from both within \cite{lancaster} and outside of economics, is still
dominant. However, recently many business practitioners started to
exploit the affinity in people's tastes
in order to predict their personal preferences and come up with
individually-tailored recommendations.
Our basic premise is that people's consumption
patterns are not based on the complete optimization over all
possible choices. Instead, they constitute just a small revealed part of the vast pool
of ``hidden wants''. These hidden wants, if properly exploited,
can lead to much better matches between people and products,
services, or other people. In the economy of the past such
opportunities were hardly exploitable. Things have changed in the course of the current
information revolution, which both connected people on an
unprecedented scale, and allowed for easy collection of the
vast amount of information on customer's preferences.
In just a few years the internet has already changed much of our traditional
perceptions about human interactions, both commercial and social.
We believe that technical advances in wireless and other network interfaces are imminent
of being able to capture the necessary information virtually free
and to put this theory to use.

Our aim is to predict yet unknown individual consumer preferences,
based on the pattern of their correlations with already
known ones. Predictive power obviously
depends on the ratio between the known and yet unknown
parts. When the fraction of known opinions $p$ is too small,
only occasional predictions are possible. When it surpasses the first threshold,
that we refer to as $p_1$, almost all unobserved preferences
acquire some degree of predictability. Finally, for $p$ above the second higher
threshold $p_2$, all these unobserved preferences can be uniquely reconstructed.
In what follows we describe a simple model of how customer's opinions are
formed and spell out in some details basic algorithms allowing for their
prediction.

To make this discussion somewhat less abstract let us consider
a {\it matchmaker} or an advisor service which already exists on many
book-selling websites that personally recommends new books to
each of their customers. In order for such recommendation to be successful
one needs to assume the existence of some ``hidden metrics'' in the space
of reader's tastes and book's features. In other words, the matchmaking
is possible only if
opinions of two people with similar tastes on two books with
similar features are usually not too far from each other. In this
work we use the simplest realization of this hidden metrics.
We assume that each reader is
characterized by an $M$-dimensional array ${\bf r}=
(r^{(1)},r^{(2)},\ldots,r^{(M)})$ of his/her tastes in books,
while each book has the corresponding list of $M$ basic
``features'' ${\bf b}=(b^{(1)},b^{(2)},\ldots,b^{(M)})$ \cite{lancaster}.
An opinion of a reader on a book is given simply by
an overlap (scalar product) $\Omega$ of reader's vector of tastes,
and book's vector of features:
$\Omega={\bf r} \cdot {\bf b}=\sum_{\alpha=1}^{M}
r^{(\alpha)}b^{(\alpha)}$. The matchmaker has some incomplete knowledge
about opinions of his customers on the
books they have read, and he uses it to reconstruct
yet unknown opinions (overlaps) and to recommend
books to its customers.

The {\it central} position of our matchmaker with respect to its customers
makes its services dramatically different from those of the so-called
``smart agents'' \cite{sa}, whose goal is to anticipate and predict tastes of their
individual owners. Indeed, the scope of recommendations of a smart agent
is severely limited by the fact that each of them serves its own master,
so that others would not cooperate. On the other hand, our matchmaker
is a completely neutral player in an economic game, who is able to synergistically
use the knowledge collected by all players/agents to everybody's advantage (including
his own).

The information about who-read-what is best
visualized as a bipartite undirected graph in which vertices
corresponding to readers are connected by edges to vertices
corresponding to books each of them has read and reported opinion to the
matchmaker. Similar graphs (or networks) were recently drawn to
the center of attention of the statistical physics community
\cite{small_world,barabasi,newman} under a name ``small
world networks''. For example, statistical properties of a
bipartite graph of movie actors connected to films they appeared
in were studied in \cite{small_world,barabasi}, while that of scientists and
papers which they co-authored - in \cite{newman}.
In this paper we go beyond empirical studies or
simple growth models of such graphs. The new feature making the graphs
introduced in this work richer than ordinary undirected
graphs is that in our graphs each vertex has a set of $M$
``hidden'' internal degrees of freedom. Consequently, each edge
carries a real number $\Omega$, representing the similarity or
overlap between these internal degrees of freedom on two vertices
it connects. In our case this number quantifies the matchmaker's
knowledge of an opinion that a given customer has on a given product.
Therefore, we would refer to such graphs as {\it knowledge or opinion
networks}.

In the most general case any two vertices in the knowledge network
can be connected by an edge. It is realized for instance if vectors
${\bf r}_1, {\bf r}_2,\ldots {\bf r}_N$ stand for strings of individual
``interests'' in a group of $N$ people. The overlap
$\Omega_{ij}={\bf r}_i \cdot {\bf r}_j$ measures the similarity of
interests for a given pair of people and can be thought of as the
``quality of the match'' between them.
The matchmaker's goal is to analyze this information and to
recommend to a customer $i$ another customer $j$, whom he has not
met yet, and who is likely to have a large positive overlap with
his/her set of interests. Mutual opinions can be conveniently
stored in an $N \times N$ {\it symmetric} matrix of scalar
products $\widehat{\Omega}$. In the above case any element of
this matrix can be in principle ``reported'' to the
matchmaker. Different restrictions imposed on this most general
scenario describe other versions of our basic model such as: 1) An
advisor service recommending $N_b$ products to $N_r$ customers
(e.g. our model of books and readers from the introduction). In
this case the square matrix $\widehat{\Omega}$ has $N_r+N_b$ rows and
columns, while all entries known to the matchmaker are
restricted to the $N_r \times N_b$ rectangle,
corresponding to opinions of customers on products.
2) A real matchmaking service recommending $N_m$ men and $N_w$ women
to each other. Here we assume that each man and woman can
be characterized by two $M$-dimensional vectors: the first one is
the vector ${\bf q}$ of his/her own ``qualities'', while the
second one ${\bf d}$ represents the set of his/her ``desires'',
i.e. desired ideal qualities that he/she is seeking in his/her
partner. The opinion of a person $i$ on a person $j$ is then given
by a scalar product ${\bf d}_i \cdot {\bf q}_j$, while the
opposite opinion has in general a completely different value
${\bf d}_j \cdot {\bf q}_i$. The full $(2N_m+2N_w)\times (2N_m+2N_w)$
overlap matrix is still symmetric but only two small sectors,
containing $N_m \times N_w$ elements each, are accessible
to the matchmaker.

With a small modification this last scenario can be applied to a completely
different problem, namely that of physical interactions between in a set of
biological molecules such as proteins.
It is known that high specificity of such interactions is achieved
by the virtue of the ``key-and-lock'' matching of features on their surfaces.
Given the space of possible shapes of locks and keys, each molecule can be
described by two vectors ${\bf l}_i$, ${\bf k}_i$ of 0's and 1's which
determine which keys and locks are present on its surface. Provided that
the key $k^{\alpha}$ uniquely fits to the lock $l^{\alpha}$, the strength of the
interaction between these two molecules is determined by
$\Omega_{ij}={\bf k}_i \cdot {\bf l}_j+{\bf k}_j \cdot {\bf l}_i$.

In the rest of the paper we
concentrate only on the most general non-bipartite case of an $N
\times N$ matrix of overlaps of interests in a group of $N$
customers and leave other more restricted situations for future work
\cite{mz_work_in_progress}.
The matchmaker always has only partial and noisy information
about the matrix $\widehat{\Omega}$ due to
several factors: 1) First and most importantly, the matchmaker
knows only some of the opinions $\Omega_{ij}$ of his customers
on each other, which he uses to guess the rest.
2) In real life the
overlap could never be precisely measured.
In the simplest case of an extremely narrow information channel
customers report to the matchmaker only the {\it sign} of their
overlap with other customers.
One can also imagine a somewhat wider channel, where
the matchmaker asks his customers to rate their satisfaction by a
grade system, the finer the better. 3)
The loss of information due to a narrow channel between the matchmaker and
its customers can be further complicated by a random noise in reporting,
which would inevitably be present in real life situations. Indeed, we are far from
assuming that the scalar product of tastes and features completely
determines the customer satisfaction with a product, or that
similarity of interests is all that matters when two people form
an opinion about each other. One should always leave room for an
idiosyncratic reaction, which does not result from any logical
weighting of features. Our hope is that strong mutually
reinforcing correlations due to the redundance of information stored in
an idealized matrix $\widehat{\Omega}$ would manifest themselves in a large
enough group of customers even when they are masked by a substantial
amount of idiosyncratic noise. In principle all these three
sources of noise and partial information are present
simultaneously. However, in this work we will treat them
separately and restrict ourselves only to the case where the matchmaker
knows the {\it exact} values of all overlaps, reported to him.
It is easy to see how
correlations between matrix elements allow the matchmaker to
succeed in his goal of prediction of yet unknown overlaps.
For example, the known values of
$\Omega_{12}={\bf r}_1 \cdot {\bf r}_2$, and $\Omega_{23}={\bf
r}_2 \cdot {\bf r}_3$ somewhat restrict the possible mutual
orientation of vectors ${\bf r}_1$ and ${\bf r}_3$, and, therefore,
contain information about the value of the yet unknown overlap $\Omega_{23}$.
Below we will demonstrate that the predictability of an overlap between
two points that are already connected by a chain of known overlaps of length $L$
is proportional to $M^{-(L-1)/2}$ and, therefore,
exponentially decays with $L$ for $M>1$. Hence, an appreciable prediction
becomes only possible when two points are connected by exponentially
many mutually reinforcing paths.

The amount of information collected by the matchmaker on its customers can
be conveniently characterized by either the number $K$ or the
density $p=2K/N(N-1)$ of known overlaps among all $N(N-1)/2$
off-diagonal elements of the matrix.
For very small $K$ all edges of the knowledge network are disconnected
and no prediction is possible. As more and more edges are
{\it randomly} added to the network,
the chance that a new edge would join two previously connected
points, i.e the probability to form a loop in the network,
increases. It is exactly in this situation the matchmaker had some
predictive power about the value of the new overlap before it was
observed. However, this excess information
would disappear in the thermodynamic limit $N \to \infty$ until
the density of edges reaches the first threshold $p_1=1/(N-1)$.
This threshold is nothing else but a percolation transition, above
which the Giant Connected Component (GCC) appears in a random graph. For
$p>p_1$ the fraction of nodes in the GCC rapidly grows, exponentially
approaching 100\%. It means that already for a moderate ratio $p/p_1$
almost every new edge added to the graph would join two previously
connected points. This transition would also manifest
itself in the behavior of the entropy of the joint probability distribution
of unknown overlaps \cite{mz_work_in_progress}.

One has to remember though that the predictive power of
the matchmaker is exponentially small for long loops. That means that
while the typical diameter of the graph is still large, the loop correlation
is too weak to significantly bias most of the unknown overlaps. The
reliable prediction becomes possible only for much higher values
of $p$. Let us calculate $p_2$ - the point of the second phase
transition, above which the values of {\it all} unknown overlaps
are completely determined by the information contained in known ones.
Using a geometrical language at this point the knowledge network
undergoes a ``rigidity percolation'' phase transition, at which
relative orientations of vectors ${\bf r}_{i}$ become fixed.
Such transition is possible only for $N>M$ since only in
this case $\widehat{\Omega}$ contains {\it redundant} information
about components of all vectors ${\bf r_i}$.
The position of the second
phase transition $p_2$ can be determined by carefully counting the
degrees of freedom. For $N>M$ the overlap matrix
$\widehat{\Omega}$ has very special spectral properties:
it has precisely $N-M$ zero eigenvalues, while the remaining
eigenvalues are strictly positive. An easy way to demonstrate
this is to recall that the overlap matrix can be written as
$\widehat{\Omega}=\widehat{R} \widehat{R}^{\dag}$, where
$\widehat{R}$ is the $N \times M$ rectangular matrix formed by
vectors $r^{(\alpha)}_i=R_{i \alpha}$. The Singular Value
Decomposition (SVD) technique allows one to ``diagonalize''
$\widehat{R}$ ($N>M$), that is to find an $M \times M$ orthogonal
matrix $\widehat{V}$, ($\widehat{V}\widehat{V}^{\dag}=1$), an $M
\times M$ positive diagonal matrix $\widehat{D}$, and an $N
\times M$ matrix $\widehat{U}$ formed by $M$ orthonormal
$N$-dimensional vectors, such that $\widehat{R}=\widehat{U}
\widehat{D}\widehat{V}$. Now it is easy to see that
$\widehat{\Omega}=\widehat{U}\widehat{D}^2
\widehat{U}^{\dag}$ has precisely $M$ {\it positive} eigenvalues equal
to squares of the elements of the diagonal matrix
$\widehat{D}$, and $N-M$ zero eigenvalues.
The number of degrees of freedom of $\widehat{\Omega}$ is equal to the
$NM$ degrees of freedom of $R$ minus $M(M-1)/2$ of the ``gauge''
degrees of freedom of the orthogonal matrix $V$, which have no
influence on elements of $\widehat{\Omega}$. Once the number of
known elements $K$ exceeds the total number of
degrees of freedom of $\widehat{\Omega}$, the remaining unknown
elements of $\widehat{\Omega}$ can be in principle reconstructed.
Therefore, the second phase transition happens at
\beq
p_2=
{M(2N-M+1) \over N(N-1)} \simeq 2M/N \qquad .
\eeq
Here the $\simeq$ sign corresponds to the limit $N \gg
M$.

Practically however, in order to calculate the set of unknown overlaps one needs
to solve a system of nonlinear equations with a huge number of unknown variables,
which is a daunting task. To this end we came up with
a simple and efficient iterative numerical algorithm,
that uses the special spectral properties of
$\widehat{\Omega}$:
(1) Construct the initial approximation $\widehat{\Omega}_a$ to
$\widehat{\Omega}$ by substituting 0 for all its {\it unknown} elements;
(2) Diagonalize $\widehat{\Omega}_a$, and construct the matrix $\widehat{\Omega}'_a$
by keeping the $M$ largest (positive)
eigenvalues and eigenvectors of $\widehat{\Omega}_a$, while
setting the remaining $N-M$ eigenvalues to zero.
(3) Construct the new refined approximate matrix $\widehat{\Omega}_a$ by
copying all unknown elements from $\widehat{\Omega}'_a$, while resetting the
rest to their {\it exactly known values}.
(4) Go to the step (2).
As shown in Fig. 1 for $p>p_2$
$\widehat{\Omega}_a$ converges to $\widehat{\Omega}$
exponentially fast in the number of iterations $n$.
Numerical simulations also indicate that the rate of
this exponential convergence scales as $(p-p_2)^2$
above the second phase transition (see the inset in Fig. 1).

Below $p_2$ this algorithm performs rather poorly and the error may
even grow with the number of iteration steps. This is to be expected since
in this region there is more than one solution for the $\widehat{\Omega}$,
consistent with a set of constraints, imposed by $K$ known matrix
elements. While our iterative algorithm always converges to {\it one}
of such solutions, barring an unlikely accident,
this solution is far from the set of ``true'' values
of unknown matrix elements.  In this situation the best thing that a
matchmaker can do is to calculate the average value $\avg{\Omega_{pq}}$
of each unknown element in the ensemble of all matrices,
consistent with a given set of $K$ constraints.
We have succeeded
in estimating $\avg{\Omega_{pq}}$ analytically. This calculation involves rather
heavy algebra and will be reported elsewhere \cite{mz_work_in_progress}.

In the above discussion the parameter $M$ was treated as fixed and
known property of the system.
However, in real life one usually does not know a priori
the number of relevant components of an idealized vector of tastes or
features. Here we want to propose a criterion on
how to optimally choose it.  If the number of known overlaps
$K$ is small, it would be useless to try to model the matrix using
a high-dimensional space of tastes. Indeed, all the free play
allowed by a large $M$ would not give the matchmaker much of a prediction
power anyway. This leads us to a conjecture
that the optimal way for a matchmaker to select
an effective number of internal degrees of
freedom $M_{\mathrm eff}$ is to do it in such a way
that the system is balanced precisely at or
near the critical threshold $p_2$. In other words,
given $K$ and $N$ one should solve the equation
$NM_{\mathrm eff}-M_{\mathrm eff}(M_{\mathrm eff}-1)/2=K$
to find  $M_{\mathrm eff}=[N+1/2-\sqrt{(N+1/2)^2-2K}] \cong K/N$.

Finally we introduce a particularly simple analytically
tractable example of an knowledge network, where each component
$r_i^{\alpha}$ of a hidden vector ${\bf r}_i$ is {\it independently}
drawn from a normal distribution.
The joint probability distribution $P(\widehat{\Omega})$ of all
$N(N+1)/2$ elements of the (symmetric) overlap matrix
$\widehat{\Omega}$ is then
given by a multidimensional integral
$P(\widehat{\Omega})=\int \! \! \int \! \! \ldots \! \!
\int \prod_{i,\alpha}
(dr_i^{(\alpha)}/\sqrt{2 \pi}) \exp \left[-\sum_{i,\alpha}
(r_i^{(\alpha)})^2/2 \right]
\prod_{i\leq j} \delta(\Omega_{ij}-\sum_{\alpha=1}^M r_i^{(\alpha)}
r_j^{(\alpha)})$. Using the standard integral representation for the
$\delta$-function, $\delta(x)=\int_{-\infty}^{\infty} \exp (i \lambda
x)\ d\lambda /(2 \pi)$, and calculating exactly the
path integral, now quadratic in $r_i^{(\alpha)}$,
one arrives at a remarkably elegant and compact expression
\cite{wishart}:
\begin{eqnarray}
P(\widehat{\Omega})=\int\!\! \int \!\! \ldots \!\!
\int_{-\infty}^{\infty}
\prod_{i \leq j}\left({ d\lambda_{ij} \over 2 \pi}
\exp(i \lambda_{ij} \Omega_{ij}) \right)
\nonumber
\\
\det(\hat{1}+i\widehat{\Lambda})^{-{M \over 2}}
\label{p_omega}
\end{eqnarray}
The matrix $\hat{1}$ is the $N \times N$ unity matrix, while
$\widehat{\Lambda}$ is a symmetric matrix with elements $2 \lambda_{ii}$
on the diagonal and $\lambda_{ij}$ off the diagonal.
This expression is the multi-dimensional Fourier
transform of the joint probability distribution $P(\widehat{\Omega})$,
so that $\Phi(\widehat{\Lambda})=\det(\hat{1}+
i\widehat{\Lambda})^{-M/2}$ is nothing else but
the {\it generating function} of this distribution!
As usual, Taylor expansion
of the generating function in powers of $\lambda_{ij}$
around $\widehat{\Lambda}=0$ allows one to calculate any imaginable
correlation between integer powers of $\Omega_{ij}$.
It is more convenient to work with irreducible correlations,
generated by the Taylor expansion of $\phi(\widehat{\Lambda})=
\ln (\Phi(\widehat{\Lambda}))=
-(M/2)\ln(\det(\hat{1}+
i\widehat{\Lambda}))=-(M/2){\mathrm Tr}[\ln(\hat{1}+
i\widehat{\Lambda})]$.
A surprising consequence of the above exact expression for
$\phi(\widehat{\Lambda})$ is that {\it all} irreducible
correlations of matrix elements are proportional to $M$.
In particular, the expansion
$\phi(\widehat{\Lambda})=(M/2)
\sum_{L=1}^{\infty}{\mathrm Tr}[(-i \widehat{\Lambda})^L ]/L$.
allows one to calculate any correlation of the type
$\avg{\avg{\Omega_{i_1i_2} \Omega_{i_2i_3} \ldots \Omega_{i_{L-1}i_L}
\Omega_{i_Li_1}}}=M$, corresponding to a given non self-intersecting loop
on the network.
The presence of such cyclic
correlations indicates that signs of matrix elements are weakly
correlated.
Taking into account that each $|\Omega_{ij}| \sim \sqrt{M}$ and
using scaling arguments it is straightforward to demonstrate
that the predictability of one of the overlaps in the loop of length
$L$ based on the knowledge of others scales as $M^{-(L-1)/2}$.

In this letter we have described a general framework allowing one
to predict elements from the unobserved part of an knowledge network
based on the observed part. Prediction power was shown to strongly depend
on the ratio between these two parts. While our original motivation was to
model a commercial matchmaking service in the internet age,
the implications go well beyond. We would like to point out
that our general framework, developed for knowledge networks,
could be also of much importance in the field of bioinformatics, where
cross-correlations, mutual interactions, and functions of large sets
of biological entities such as proteins, DNA binding sites, etc.,
are only partially known. It is conceivable that a similar approach
applied to e.g. a large matrix of protein-protein interactions \cite{fields}
would prove to be fruitful.

We have benefitted from conversations with T. Hwa,  M.
Marsili, C. Tang, Y. Yu and A. Zee. Work at Brookhaven National Laboratory
was carried out under Contract No. DE-AC02-98CH10886, Division of
Material Science, U.S.\ Department of Energy. This work was
sponsored in part by Swiss National Foundation under Grant 20-61470.00.

\begin{figure}
\epsfxsize=3.5in
\epsffile{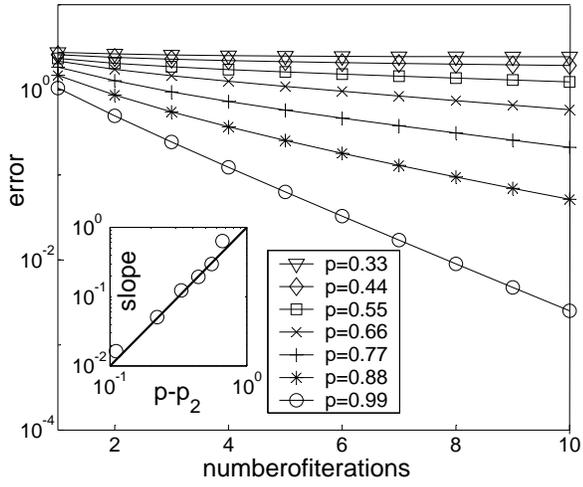}
\caption{The average error in the value of unknown matrix elements of $\widehat{\Omega}$ as a
function of the number of iterations. All $r_i^{(\alpha)}$ are independent Gaussian random numbers.
The parameters of the model are $M=9$, $N=50$, corresponding to $p_2=0.34$.
The inset shows the scaling of an exponential convergence rate as a function of $p-0.34$.
The solid line has the slope $2$. }
\end{figure}

\end{document}